\documentclass[a4paper,11pt]{article}
\pdfoutput=1 

\usepackage{jinstpub} 

\usepackage{lineno}
\usepackage{wrapfig}
\usepackage{hyperref}

\title{\boldmath Data Processing Engine (DPE): Data Analysis Tool for Particle Tracking and Mixed Radiation Field Characterization with Pixel Detectors Timepix}

\author[a,b,1]{L. Marek,\note{Corresponding author.}}
\author[a]{C. Granja,}
\author[a]{J. Jakubek,}
\author[a]{J. Ingerle,}
\author[a]{D. Turecek,}
\author[c]{M. Vuolo,}
\author[a]{C. Oancea}

\affiliation[a]{ADVACAM, U Pergamenky 12, Prague 7, Czech Republic}
\affiliation[b]{Faculty of Mathematics and Physics, Charles University
, V Holesovickach 2, Prague 8, Czech Republic}
\affiliation[c]{European Space Agency, ESTEC,
Keplerlaan 1, NL-2200 AG Noordwijk, Netherlands}
\emailAdd{lukas.marek@advacam.cz}

\abstract{
Hybrid semiconductor pixelated detectors from the Timepix family are advanced detectors for online particle tracking, offering energy measurement and precise time stamping capabilities for particles of various types and energies. This inherent capability makes them highly suitable for various applications, including imaging, medical fields such as radiotherapy and particle therapy, space-based applications aboard satellites and the International Space Station, and industrial applications. The data generated by these detectors is complex, necessitating the development and deployment of various analytical techniques to extract essential information. 
For this purpose, and to aid the Timepix user community, it was designed and developed the 
"Data Processing Engine" (DPE) as an advanced tool for data processing designed explicitly for Timepix detectors. The functionality of the DPE is structured into three distinct processing levels: i) Pre-processing: This phase involves clusterization and the application of necessary calibrations and corrections. ii) Processing: This stage includes particle classification, employing machine learning algorithms, and the recognition of radiation fields. iii) Post-processing: Involves various analyses, such as directional analysis, coincidence analysis, frame analysis, Compton directional analysis, and the generation of physics products, are performed. The core of the DPE is supported by an extensive experimental database containing calibrations and referential radiation fields of typical environments, including protons, ions, electrons, gamma rays and X rays, as well as thermal and fast neutrons. To enhance accessibility, the DPE is implemented into various user interface platforms such as a command-line tool, an application programming interface, and as a graphical user interface in the form of a web portal. The DPE's broad utility is exemplified through its integration into various applications and developments.
}


\begin{document}

\maketitle
\flushbottom

\section{Introduction and detectors}
\begin{figure}
\centering 
\includegraphics[width= 11cm]{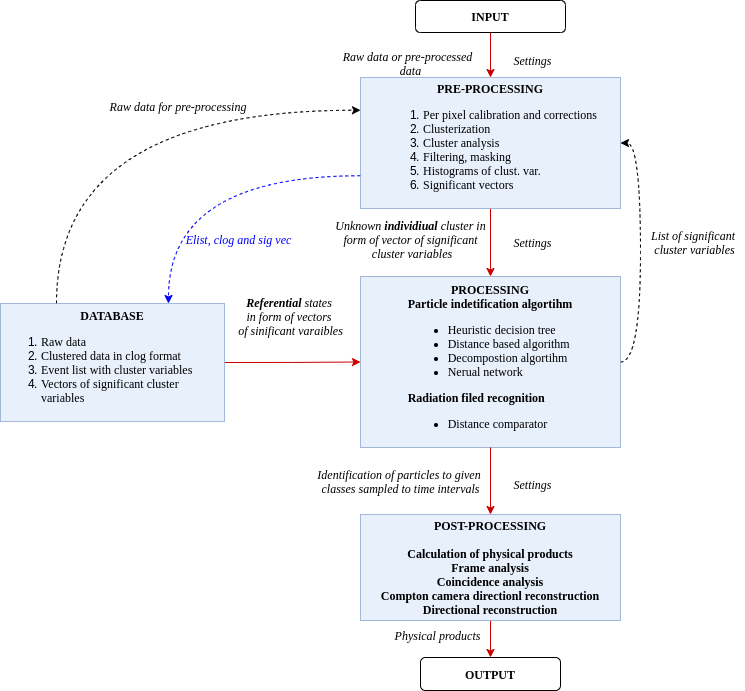}
\caption{Diagram of DPE architecture highlighting communication and functionality of its main components: pre-processing, processing, post-processing and database.}
\label{fig:digram_architecture}
\end{figure}

The Timepix family of hybrid semiconductor pixelated detectors, including Timepix1 (TPX), Timepix3 (TPX3), and Timepix2 (TPX2), has proven to be a reliable tool for radiation spectrometry and particle tracking~\cite{HEI13,BAL18}. These detectors offer the unique advantage of precise energy measurements and time-stamping capabilities, making them highly adaptable to a wide range of scientific and industrial applications such as space radiation monitoring \cite{4}, particle radiotherapy \cite{1,3}, imaging with X-rays and particles \cite{8, 9}, Compton camera, radiation tracking \cite{5}, dosimetry \cite{7}, neutron detection \cite{1,2}, etc.
Despite their vast potential, the complex data generated by Timepix detectors necessitates the development of advanced data processing techniques to explore their full potential. To address this challenge, we designed and developed the "Data Processing Engine" (DPE), a comprehensive software tool developed specifically for Timepix detectors. DPE addresses the need for efficient data analysis and interpretation with main functional elements as illustrated in Figure \ref{fig:digram_architecture}. The development was based on a set of fundamental principles, aimed at addressing the complex data analysis for Timepix family detectors of various sensor materials and thicknesses (Si, CdTe, GaAs, SiC) \cite{13}: i) consolidation of complex data analysis under one engine (particle type identification, tracking, Compton camera, etc.), ii) creation of a database with reference data (calibration and mixed field data), iii) development of robust and verified tool, iv) creation of user-friendly platform as the web portal, v) accessibility also for non-experts. These capabilities make DPE suitable for various use cases, including space exploration, medical applications, environmental studies, and educational purposes.

\section{Pre-processing} 
\begin{figure} 
\centering 
\includegraphics[width= 15cm]{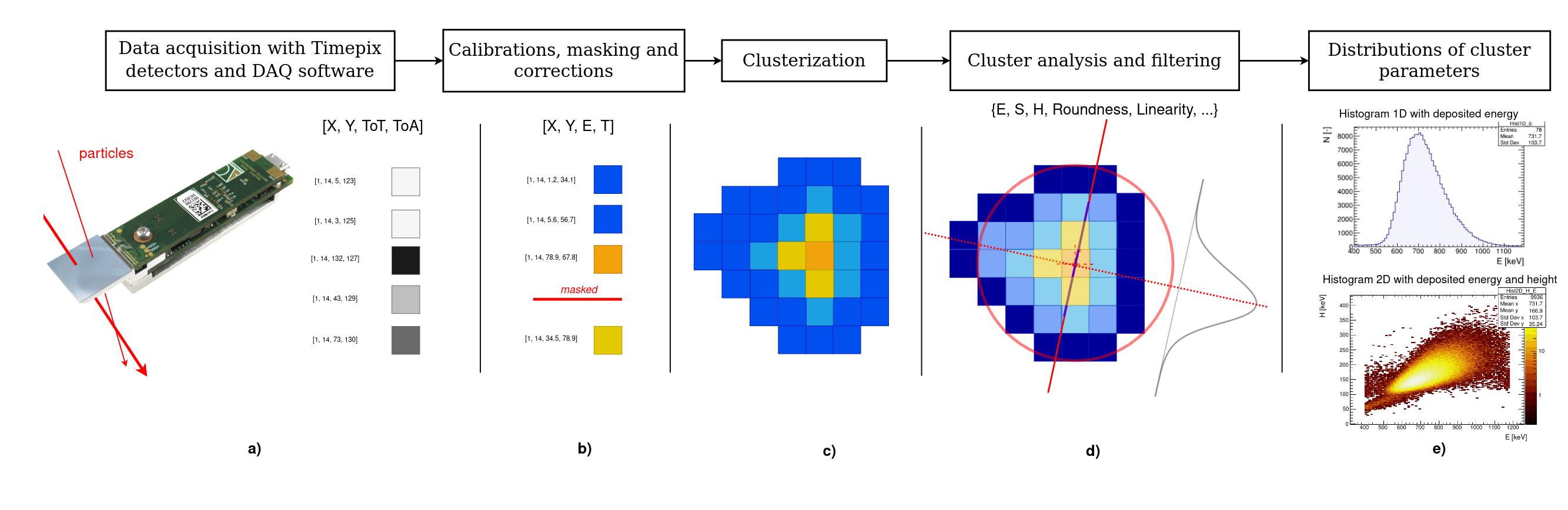}
\caption{Pre-processing stage flow: (a) data are collected using a Timepix-based detector and raw data display ToT and ToA of particles which is in (b) converted into deposited energy and time of arrival and (c) single particles are identified during clusterization. (d) The cluster is analyzed and the parameters of each particle are determined and further used in (e) to provide distributions of cluster parameters.}
\label{fig:pre_proc_flow}
\end{figure}

The purpose of this stage is to transform pixelated data into sets of correlated pixels, clusters, an application of calibration and corrections, cluster analysis - performed with high-resolution pattern-recognition algorithms~\cite{HOL08}, a filtering and creation of distributions of cluster parameters. 
A diagram illustrating the segments for the pre-processing stage can be seen in Figure \ref{fig:pre_proc_flow}.
DPE is designed to accept and process data in various formats produced primarily by the PIXET SW tool~\cite{PIXET}. 

\subsection{Clusterization, calibration, masking and corrections} 
The degree of correlation depends on the type of detector employed for data acquisition. When using the TPX detector, only spatial correlation is available and subsequently utilized in the clustering process. In contrast, the TPX3 detector provides both spatial and temporal information. This additional temporal data enables a more precise clustering process, avoiding unwanted effects such as pile-up. During this stage, a fundamental energy calibration is applied to convert digital Time-over-Threshold (ToT) information into energy units, specifically in kiloelectronvolts (keV). This calibration process is accompanied by supplementary high-energy calibrations corrections \cite{15} aimed to maximally suppress the detector's unwanted or anomalous behavior. The following list includes the most important ones: high energy calibration/correction, overshoot, Volcano effect, cluster size energy correction, pile-up recognition, halo effect correction, and correction for XRF peak suppression. 

\subsection{Cluster analysis, filtering and distributions}
\begin{figure} 
\centering 
\includegraphics[width= 15cm]{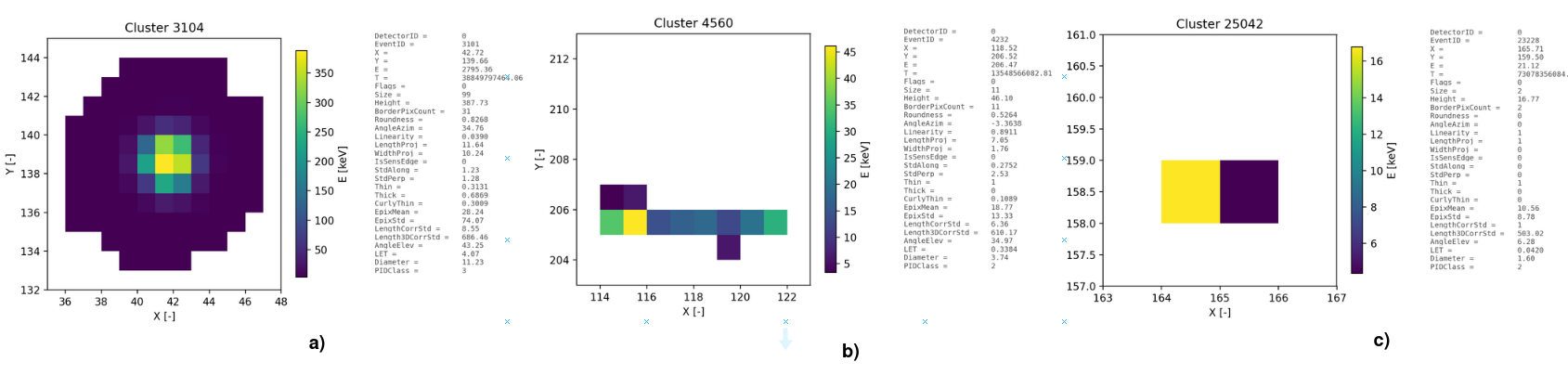}
\caption{Example of clusters and their parameters made by (a) ions, (b) electrons, and (c) X-rays measured by a Timepix detector with a silicon sensor of 500 $\mu m$ thickness. Cluster parameters can be seen for all clusters in text format next to the graphical visualization.}
\label{fig:clusters_vis}
\end{figure}

\begin{figure}
\centering 
\includegraphics[width= 14cm]{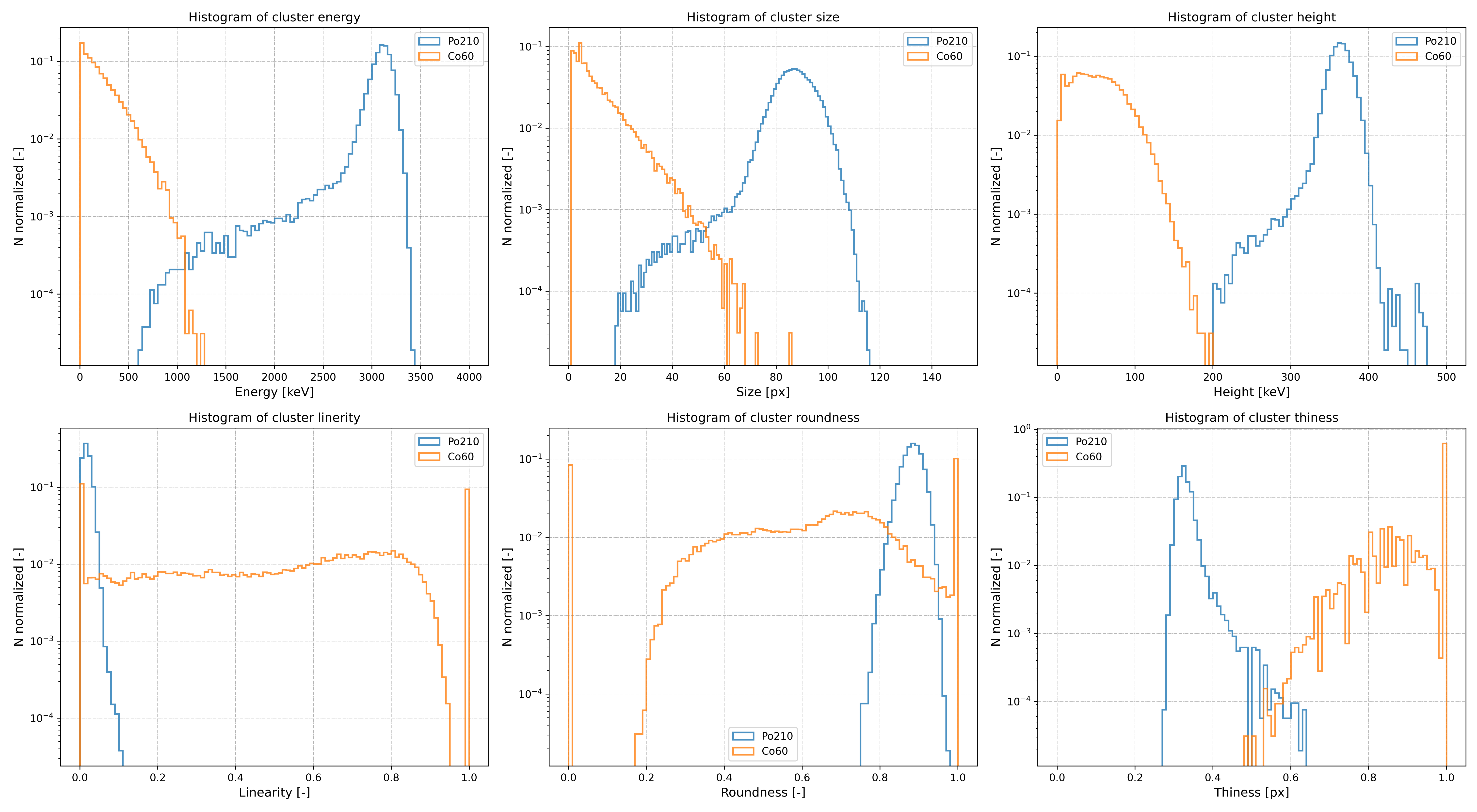}
\caption{Histogram of cluster parameters: (a) deposited energy, (b) size, (c) height, (d) linearity, (e) roundness, and (f) thinness of tracks measured by Minipix TPX3 silicon sensor of 500 $\mu$m thickness. The data was acquired from two radiation sources: $^{210}$Po (blue, mainly ions) and $^{60}$Co (orange, mainly photons and scattered electrons from the Compton effect).}
\label{fig:hist}
\end{figure}

Single particle tracking\cite{BER20} and track morphology analyses~\cite{HOL08} provide characteristic morphological and spectral parameters which are used for particle-type recognition and discrimination~\cite{GRA18}. This is illustrated in Figure \ref{fig:clusters_vis} where three clusters produced by different particles are shown together with selected feature values. Subsequent filtering can be performed to suppress unwanted events. Histograms of final clusters are created for all parameters as seen in Figure \ref{fig:hist} (selected parameters: deposited energy, size, height, linearity, roundness, and thinness of tracks, see description of parameters in \cite{6}).

\subsection{Significant vectors}
The histograms are used for radiation field recognition based on exploiting so-called significant vectors. These vectors  are compared against a reference database of known vectors associated with various radiation fields. The primary objective of this process is to determine the closest match to the database vectors, thereby estimating the type of radiation field (more detailed information will be provided in the next stage).

\section{Processing}
The primary objective of this processing stage is to perform two essential data analyses: i) Particle Identification/Classification and ii) Radiation Field Recognition.
\subsection{Particle classification}

\begin{figure}
\centering 
\includegraphics[width= 11cm]{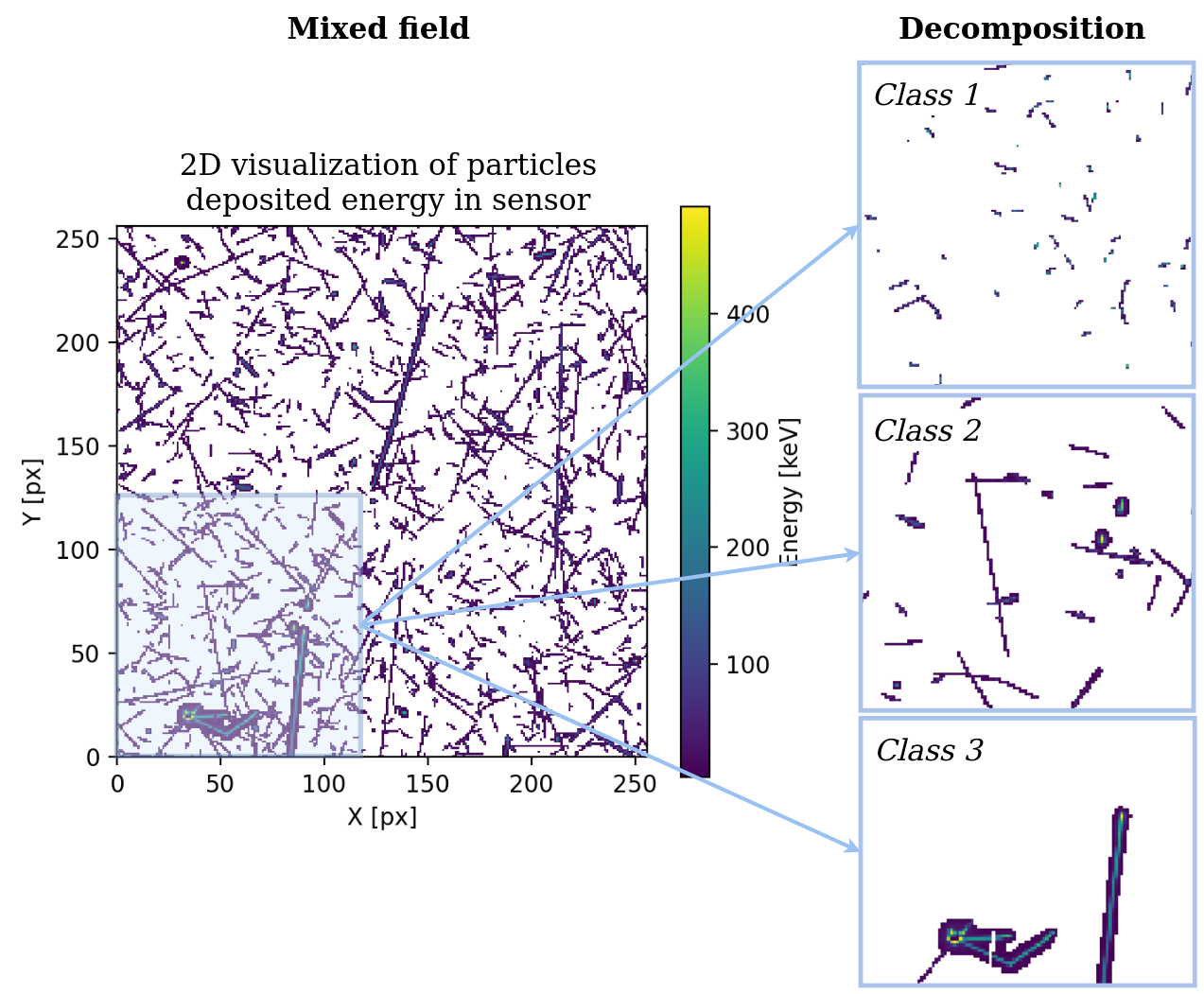}
\caption{Integrated detection and 2D visualization of deposited energy in the TPX3 detector placed in a mixed field. (left) The whole sensor pixel matrix is shown (256×256 pixels). 
(right) Enhanced visualization of a smaller area of the sensor (125×125 pixels) with field decomposition and identification using the neural network machine learning algorithms. Three classes of particles from top to bottom: Class 1 (electrons \& photons), Class 2 (protons), and Class 3 (ions) were identified.}
\label{fig:pid}
\end{figure}

The particle classification aims to provide a separation of clusters into several classes based on their morphological and spectral parameters. In the optimal case, these classes would be in accordance with particle species (electron, proton, ions, etc.) combined with their kinetic energies and directions. An example of particle classification exploiting machine learning algorithms (neural network) can be seen in Figure \ref{fig:pid}. The individual particles are separated into three classes: Class 1 called protons, Class 2 called electrons and photons, and Class 3 called ions. The discrepancy between ideal PID and presented classification can be observed in Class 2 where highly energetic protons might be also present with a certain probability. The final interpretation has to be always given by the user knowing the efficiency of the algorithm. \\
Depending on the specific objectives of data processing, such as estimating particle Linear Energy Transfer (LET) for radiotherapy \cite{5,6}, users may need to perform different types of classifications. To address this need, various algorithms have been developed, employing machine learning, heuristic, and analytic approaches: i) Heuristic decision tree, ii) Distance/similarity algorithm, and iii) Neural networks. The algorithms were designed for two main detector configurations:
i) TPX silicon 300 $\mu m$ and ii) TPX3 silicon 500 $\mu m$. \\
An additional perspective arises when considering particle classification with Timepix detectors. This perspective involves two distinct approaches: individual and statistical. The investigation of individual particles can be very effective from the classification and directional point of view. In some cases when these tracks can be very minimal concerning the size and afterward the overall information, studying individual tracks is not sufficiently rigorous for classification. This can be complemented with the investigation of overall measured information of the whole radiation field, so-called statistical methods, or radiation field recognition.

\subsection{Radiation Field Recognition}

The second important analysis in the processing stage is the radiation field recognition. Its main scope is to estimate the radiation field type based on similarity with known fields and take into account the whole measured information. To achieve this comparison in some reasonable data frame, additional abstraction of so-called significant vectors was introduced. These objects, arrays/vectors of numbers, should uniquely describe the given radiation field. Having premeasured several radiation fields, a comparison of these vectors can be done and the biggest accordance can be investigated which is then marked as the type of the radiation field.

\section{Post-processing}

\subsection{Physics products}
\begin{figure}
\centering 
\includegraphics[width= 14cm]{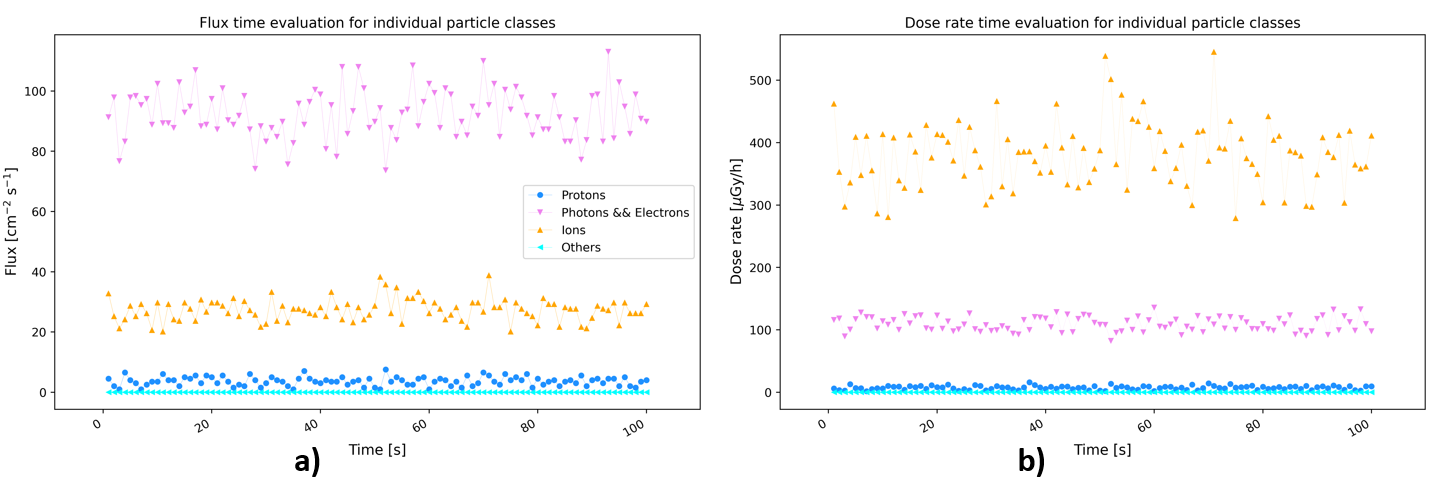}
\label{fig:flux_dr}
\caption{Time plot of particle (a) flux of particles and (c) dose-rate for a measurement in a mixed field where various classes of particles were identified using classification with the neural networks: Class 1 (protons), Class 2 (electrons and photons), Class 3 (ions), and Class 4 (other particles). The measured radiation field is a mixture of $^{210}$Po and $^{60}$Co.}
\end{figure}

The final part of processing exploits results given by previous stages and converts them into physics products such as: Count of particles, Count rate, Fluence, Flux, Dose and Dose rate. These quantities are evaluated for all particles, and an example of these products is shown in \ref{fig:flux_dr}.

\subsection{Directional and Compton analysis}
\begin{figure}
\centering 
\includegraphics[width= 14cm]{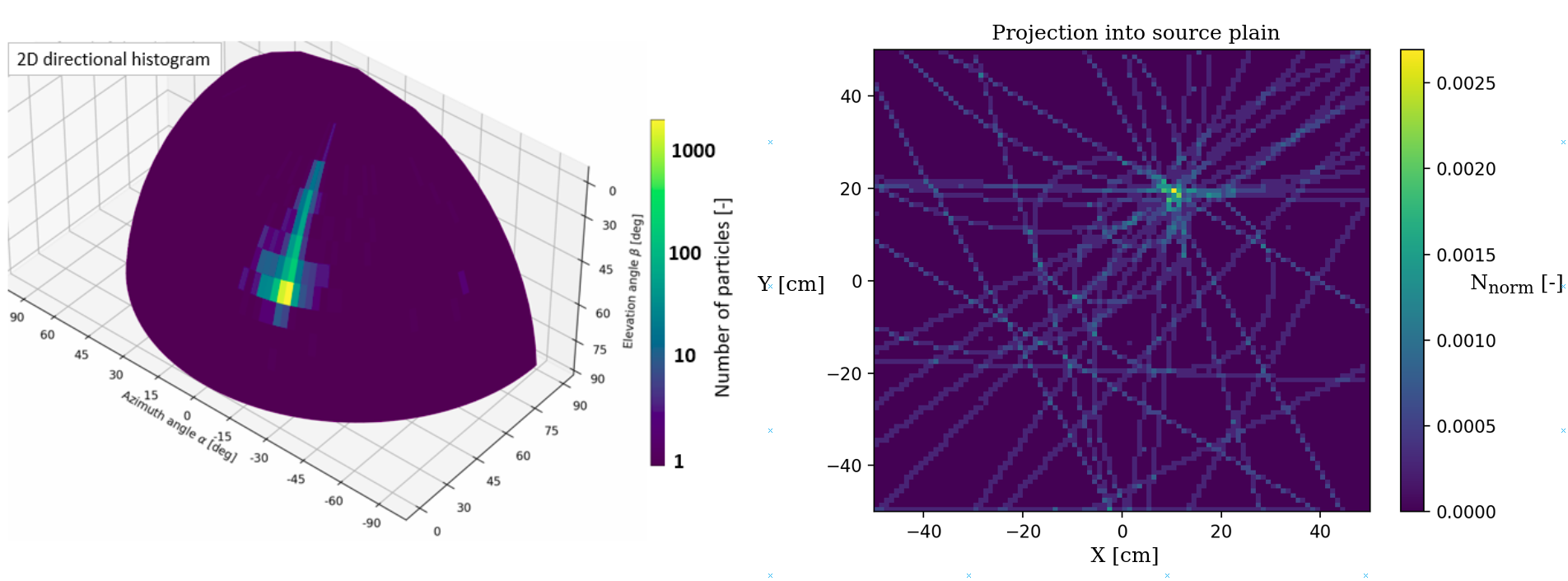}
\caption{(left) Directional 2D map of incident direction of detected particles. Highly energetic protons impacting the detector under an elevation angle of approximately 60 degrees were used. (right) Example of Comptom camera directional reconstruction of a $^{60}$Co gamma-ray source. An estimation of the source position is made at X and Y coordinates in the source plane. The source was placed at X = 10 cm and Y = 20 cm.}
\label{fig:direction_compton}
\end{figure}

The scope of this analysis is to exploit the cluster morphology and estimate the particle direction in the sense of azimuth and elevation angle. The azimuth angle can be calculated based on the approximation of the cluster as a rigid plane body. Calculating and minimization of the momentum of inertia is the core of the final estimation. The elevation angle is in the favorable cases estimated based on the geometrical idea of particle full cross complemented with correction for charge sharing (length in the sensor plane deriving from polar angle is corrected for charge sharing effect which smears the true track length of the particle) \cite{12}. This analysis has its limitations, for example in the cases when a particle is fully stopped in the sensor or the incident/elevation angle of a particle is close to the perpendicular direction to the sensor. These conditions usually result in estimation burdened with large errors. An example of the results can be seen in the 3D spatial map shown in Figure \ref{fig:direction_compton} (left).   \\
The Compton camera directional reconstruction exploits coincidences of events/photons which most likely interacted with the detector in the area of the Compton effect and afterward, they were absorbed in the sensor (via photo effect). Based on the physics perspective, if these conditions are fulfilled then information about direction can be derived from the measured data, an example can be seen in Figure \ref{fig:direction_compton} (right).

\subsection{Coincidence and Frame analysis}
The coincidence analysis aims to provide an additional framework for the investigation of coincidence events. Several clusters are marked as coincidences if they are closely time-related (in order of hundreds of nanoseconds).\\
The frame analysis main purpose is to supply information about statistical features and physics products of the frames. One of the most simplistic and very important measurement-related information is about the occupancy of the frame which can be directly exploited during the measurement using DPE to potentially avoid a loss of the data integrity. Another use of this feature is for integrated deposited energy and number of hit pixels \cite{10}. 

\section{Architecture and user interfaces}
\label{sec:platforms}

The DPE is a versatile software tool built on a C++ architecture, optimized for fast data processing. Its modular design facilitates easy integration of functionality extensions. It is created for Linux and Windows operation systems, with a Docker support. The DPE offers a command-line tool for scripting and easy first access, a well-documented API for Python and C++ programming integration, a web portal for graphical access and server processing, and a desktop application for offline analysis with parameters that can be easily fine-tuned. The DPE can accessed on the following link:  \href{https://wiki.advacam.cz/wiki/DPE}{https://wiki.advacam.cz/wiki/DPE}.


\section{Conclusions}
\label{sec:conclusion}

This paper introduces and describes the Data Processing Engine (DPE): Data Analysis Tool for Particle Tracking and Mixed Radiation Field Characterization with Pixel Detectors Timepix. It is a specialized software tool developed for Timepix family detectors by ADVACAM. The DPE addresses the complex data generated by these detectors and offers a structured approach to data processing, including pre-processing, processing, and post-processing stages. The pre-processing stage involves data calibration, and corrections for various scenarios of incomplete data collection or high-energy correction. Clusterization, cluster analysis and histograms of cluster parameters contribute to the extraction of significant vectors used in radiation field recognition. In the processing stage, the DPE performs particle classification, identifying particle types and energies. It also recognizes radiation fields, providing valuable information for various applications such as high-intensity fluxes in space or in particle therapy. The post-processing phase converts data into physics products, such as particle counts, dose rates, and flux, contributing to a deeper understanding of the data. Directional analysis, Compton analysis, coincidence analysis, and frame analysis further enhance data interpretation. An important part of the DPE is a database containing referential radiation fields of typical environments. The tool is accessible in several platforms: command-line tool, API and web portal to address different needs of user's analyses. 
Its comprehensive approach and user-friendly design make it a valuable tool for a wide range of scientific and industrial applications, from space exploration to medical fields, education, radiation dosimetry, etc.

\appendix

\acknowledgments
Work at ADVACAM was performed in frame of Contract No. 4000130480/20/NL/GLC/hh from the European Space Agency.

\end{document}